\documentclass[12pt]{article}
\usepackage{graphicx}
\usepackage[utf8]{inputenc}
\usepackage[numbers,sort&compress]{natbib} 
\usepackage{hyperref}

\newcommand\pubnumber{}
\newcommand\pubdate{\today}

\def\Title#1{\begin{center} {\Large #1 } \end{center}}
\def\Author#1{\begin{center}{ \sc #1} \end{center}}

\newcommand\pubblock{\rightline{\begin{tabular}{l} \pubnumber\\
         \pubdate  \end{tabular}}}
\newenvironment{Abstract}{\begin{quotation}  }{\end{quotation}}
\newenvironment{Presented}{\begin{quotation} \begin{center} 
             PRESENTED AT\end{center}\bigskip 
      \begin{center}\begin{large}}{\end{large}\end{center} \end{quotation}}
\def\Acknowledgements{\bigskip  \bigskip \begin{center} \begin{large}
             \bf ACKNOWLEDGEMENTS \end{large}\end{center}}

\def\beq{\begin{equation}}
\def\eeq#1{\label{#1}\end{equation}}
\def\eeqn{\end{equation}}

\def\beqa{\begin{eqnarray}}
\def\eeqa#1{\label{#1}\end{eqnarray}}
\def\eeqan{\end{eqnarray}}

\let\bar=\overbar

\def\Dslash{\not{\hbox{\kern-4pt $D$}}}
\def\dslash{\not{\hbox{\kern-2pt $\del$}}}

\def\msb{{\bar{\ssstyle M \kern -1pt S}}}

\renewcommand*{\thefootnote}{\fnsymbol{footnote}}

\begin{document}
\begin{titlepage}
\pubblock

\vfill
\Title{Probing the CP nature of the Higgs coupling in $t\bar th$ events at the LHC}
\vfill
\Author{
E. Gouveia$^1$\footnote{Funded by Funda\c{c}\~{a}o para a Ci\^{e}ncia e Tecnologia, through grant PD/BD/128231/2016}
\\[2mm]
{\rm also on behalf of}
\\[2mm]
S.P. Amor dos Santos$^2$,
M.C.N. Fiolhais$^{2,4}$,
R. Frederix$^5$, 
R. Gon\c{c}alo$^3$, 
R. Martins$^{1}$, 
A. Onofre$^{1}$,
C.M.Pease$^{4}$,
H. Peixoto $^6$,
\mbox{A. Reigoto$^1$,} 
\mbox{R. Santos$^{7,8}$,} 
J. Silva$^6$
\\[3mm]
{\footnotesize {\it 
$^1$ LIP, Departamento de F\'{\i}sica, Universidade do Minho, 4710-057 Braga, Portugal\\
$^2$ LIP, Departamento de F\'{\i}sica, Universidade de Coimbra, 3004-516 Coimbra, Portugal\\
$^3$ LIP, Av. Prof. Gama Pinto, n.2, 1649-003 Lisboa, Portugal;\\ Faculdade de Ci\^{e}ncias da Universidade de Lisboa, Campo Grande, 1749-016 Lisboa, Portugal\\
$^4$ Science Department, Borough of Manhattan Community College, City University of New York, 199 Chambers St, New York, NY 10007, USA\\
$^5$ Technische Universit\"{a}t M\"{u}nchen, James-Franck-Str.~1, D-85748 Garching, Germany \\
$^6$ Centro de F\'{\i}sica, Universidade do Minho, Campus de Gualtar, 4710-057 Braga, Portugal\\ 
$^7$ Instituto Superior de Engenharia de Lisboa - ISEL, 1959-007 Lisboa, Portugal\\
$^8$ Centro de F\'{\i}sica Te\'{o}rica e Computacional, Faculdade de Ci\^{e}ncias, Universidade de Lisboa, Campo Grande, Edif\'{\i}cio C8 1749-016 Lisboa, Portugal}}
}
\vfill
\begin{Abstract}
The determination of the $CP$ nature of the top quark Yukawa coupling is addressed in this phenomenological work, using $t\bar th$ events at the \linebreak \mbox{$\sqrt{s}=13$ TeV} LHC, with $h\rightarrow b\bar b$ and through the dilepton decay channel of $t\bar t$. Signal events were generated for pure $CP$-even, pure $CP$-odd and mixed-$CP$ coupling scenarios. Standard Model (SM) backgrounds were also considered. Angular distributions and asymmetries were studied and found to be potentially good probes to the $CP$ nature of the coupling, even after detector simulation and full event reconstruction using a kinematic fit. Expected limits to the $t\bar th$ production cross-section times branching ratio were obtained for a range of top quark Yukawa $CP$-mixing angles and for different integrated luminosities of the LHC.
\end{Abstract}
\vfill
\begin{Presented}
TOP2017 - 10th International Workshop on Top Quark Physics\\
Braga, Portugal, September 17--22, 2017
\end{Presented}
\vfill
\end{titlepage}
\def\thefootnote{\fnsymbol{footnote}}
\setcounter{footnote}{0}

\section{Introduction}

The $CP$ nature of the couplings to gauge bosons of the recently discovered particle compatible with the 125 GeV Higgs boson \cite{higgsdiscoveryatlas,higgsdiscoverycms} has been studied by ATLAS and CMS. 
So far, no equivalent analyses have been carried out in the fermionic sector alone. The top quark, due to its expected large Yukawa coupling, will be the obvious candidate for such measurements. The top quark Yukawa interaction with the Higgs boson can be parametrised as
\begin{equation}
 \mathcal{L}=\kappa y_t \bar t (\cos\alpha+i\gamma_5\sin\alpha)th,
\end{equation}
where the SM is recovered for $\alpha=0$ and $\kappa=1$, and $\alpha\notin \{0,\pi\}$ implies a non-zero $CP$-odd component in the coupling.
The Higgs associated production process $t\bar th$ 
allows a direct measurement of this coupling. While a non-zero $CP$-odd component is expected to affect the inclusive $t\bar t h$ production cross-section, it has been suggested \cite{madspin,rohini2,gunion,demartin} that differential cross-section measurements, exploiting the distinct kinematics and spin correlations among intermediate and final-state particles, can enhance sensitivity to its presence. We show that angular distributions of the kind introduced in \cite{angdist} are sensitive to the $CP$ nature of the coupling. Also, a survey of previously proposed observables resulted in a reduced set of those which still show significant discrimination power even after detector simulation, event selection and event reconstruction via a kinematic fit \cite{cpangvars}.

\section{Event generation, simulation and reconstruction}

Events from $pp$ collisions were generated at $\sqrt{s}=13$ TeV using \linebreak {\sc MadGraph5\_aMC@NLO}~\cite{madgraph}. Signal $t\bar t h$ events, as well as the irreducible background $t\bar t b \bar b$, were generated at NLO in QCD. 
Signal samples in which the Higgs has a non-zero $CP$-odd component were generated using the \texttt{HC\_NLO\_X0} UFO model~\cite{demartin}, with $\cos\alpha$ varying from -1 to 1. Particle decays were handled by {\sc MadSpin}~\cite{madspin}. The SM backgrounds $t\bar t+$jets, $t\bar t+V(V=Z,W)$, single top, $V+$jets, $V+b\bar b$ and $VV$ were also generated. 
Parton shower and hadronisation were performed with {\sc Pythia6}~\cite{pythia} and followed by a fast detector simulation with the {\sc Delphes}~\cite{delphes} package.

Events were selected with at least four jets (of which at least three must be $b$-tagged) and exactly two opposite-charge leptons, all with $p_T\geq20$ GeV and \mbox{$|\eta|<2.5$}. Additionally, the dilepton mass $m_{\ell\ell}$ must meet the requirement \linebreak \mbox{$|m_{\ell\ell}-91\textrm{ GeV}|>10\textrm{ GeV}$}. Kinematic reconstruction of the event starts with the choice of the most likely assignment of jets to $b$-quarks from $t$, $\bar t$ and $h$ decay, using a boosted decision tree (BDT). 
Then, a solution for the neutrino momenta is obtained by combining information from missing $E_T$ with a constraint on $W$ and $t$ masses.

\section{Angular distributions}

\begin{figure}[htb]
\begin{center}
\begin{tabular}{ccc}
\includegraphics[width=0.45\textwidth]{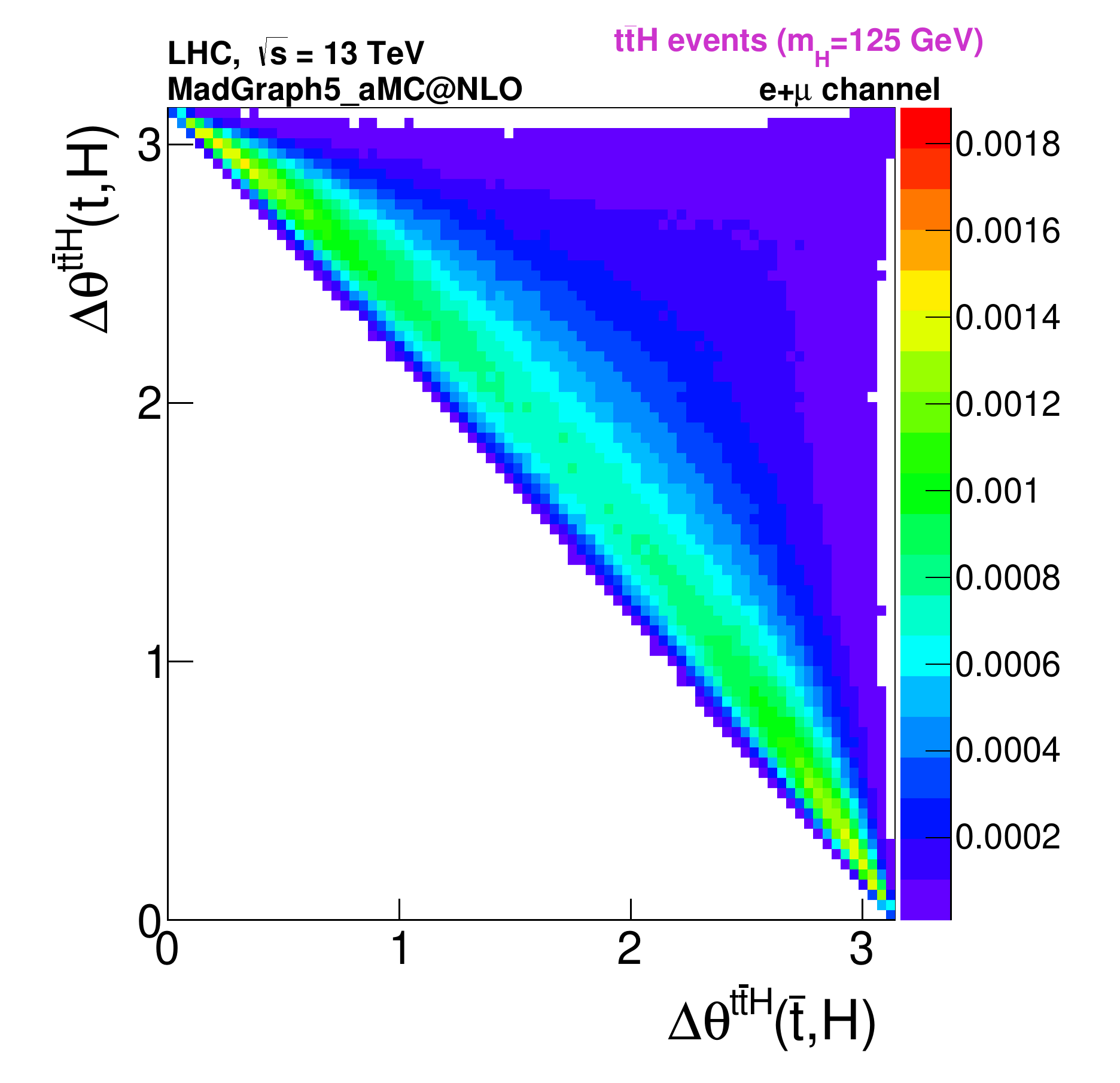} & \quad & 
\includegraphics[width=0.45\textwidth]{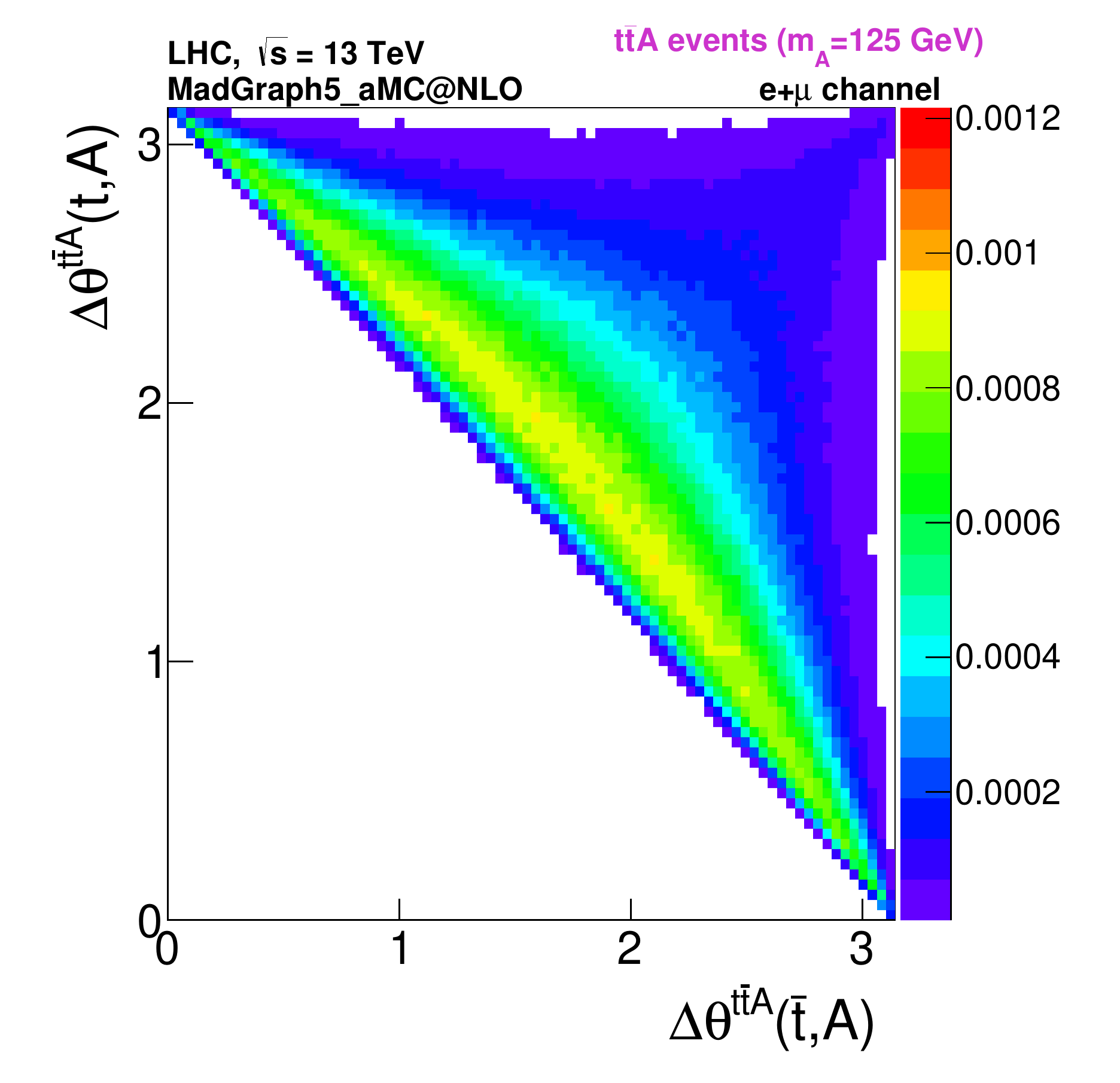}\\[-4mm]
\end{tabular}
\caption{Parton-level distributions of $\Delta\theta^{t\bar th}(\bar t,h)$ ($x$-axis) and $\Delta\theta^{t\bar th}(t,h)$ ($y$-axis). The SM signal ($h=H$) distribution (left) and the pure $CP$-odd scenario signal ($h=A$) distribution (right) are shown.}
\label{ttHangles}
\end{center}
\end{figure}

Let $\Delta\theta^{t\bar th}(t,h)$ (respectively $\Delta\theta^{t\bar th}(\bar t,h)$) be the angle between the direction of $h$ and the direction of $t$ (respectively $\bar t$), both evaluated in the $t\bar t h$ rest-frame. Figure \ref{ttHangles} shows two-dimensional distributions of signal events for these angles, calculated at NLO after parton-shower effects, prior to detector simulation and event selection.
In the SM scenario (left) the Higgs boson tends to travel preferentially along a direction close to that of one of the top quarks and nearly opposite to the other. In the pure $CP$-odd scenario (right), events are more evenly spread over the allowed range. 

\begin{figure}
\begin{center}
\begin{tabular}{ccc}
\includegraphics[width=0.45\textwidth]{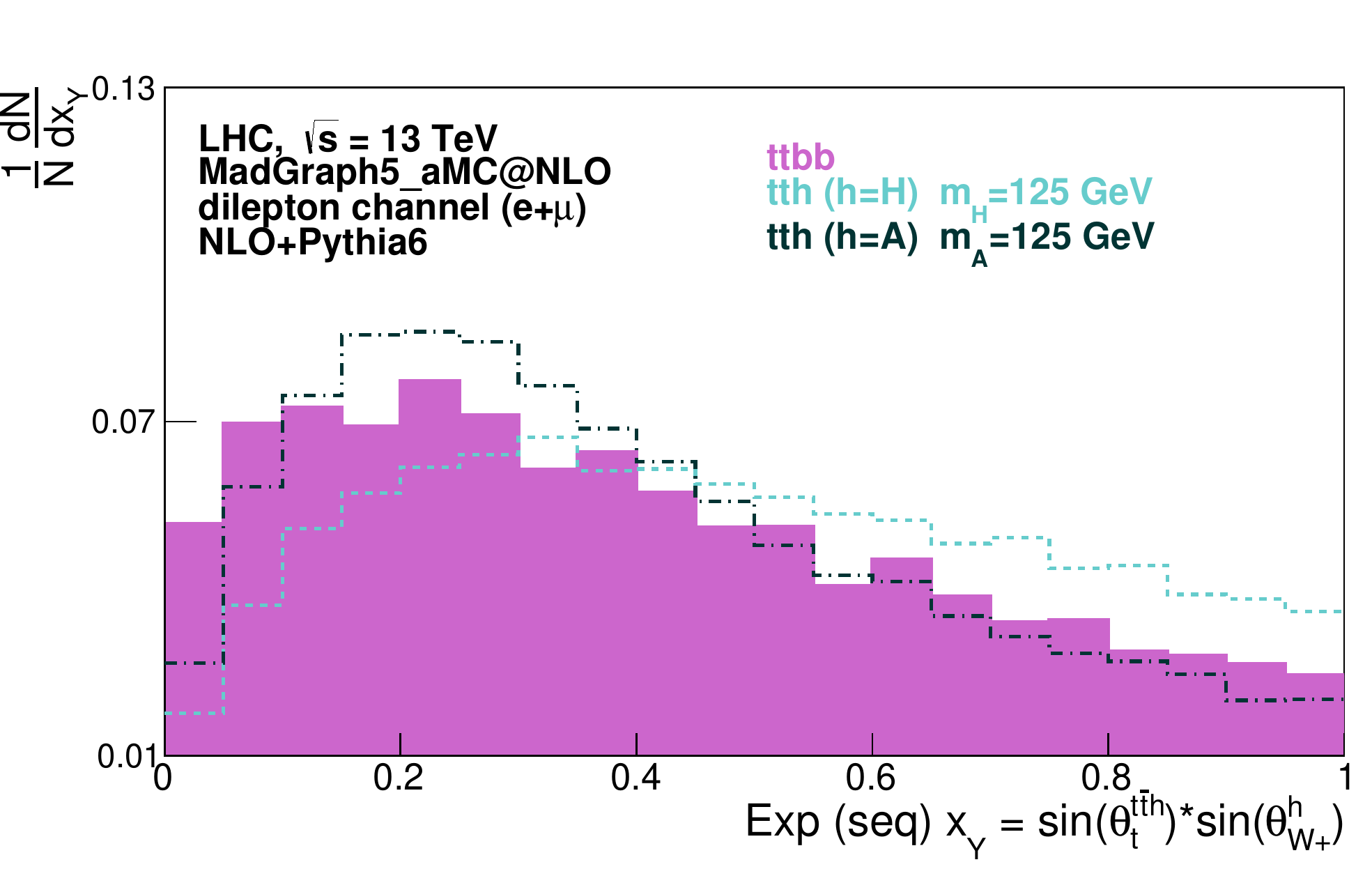} & \quad & 
\includegraphics[width=0.45\textwidth]{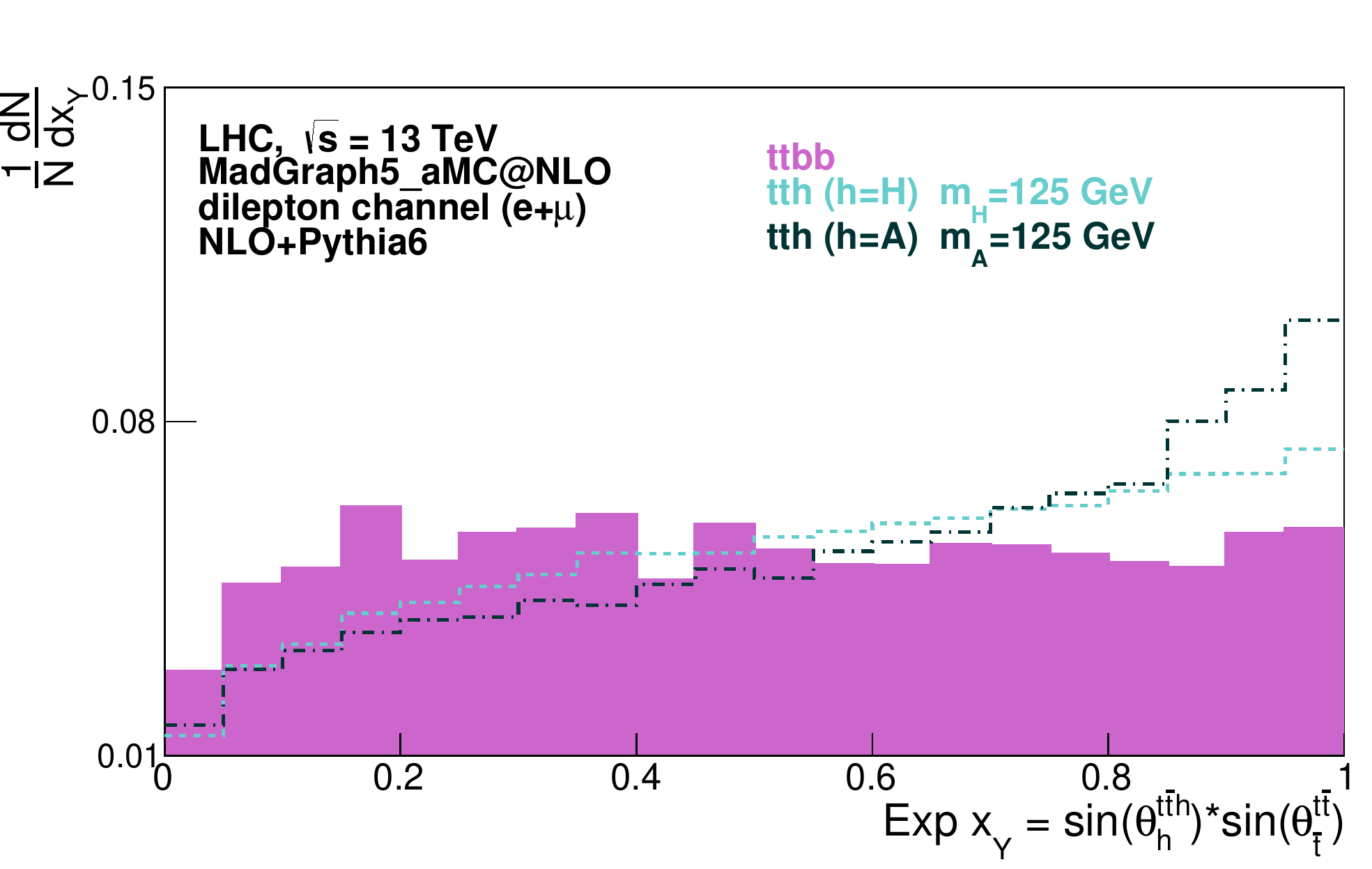}\\[-2mm]
\includegraphics[width=0.45\textwidth]{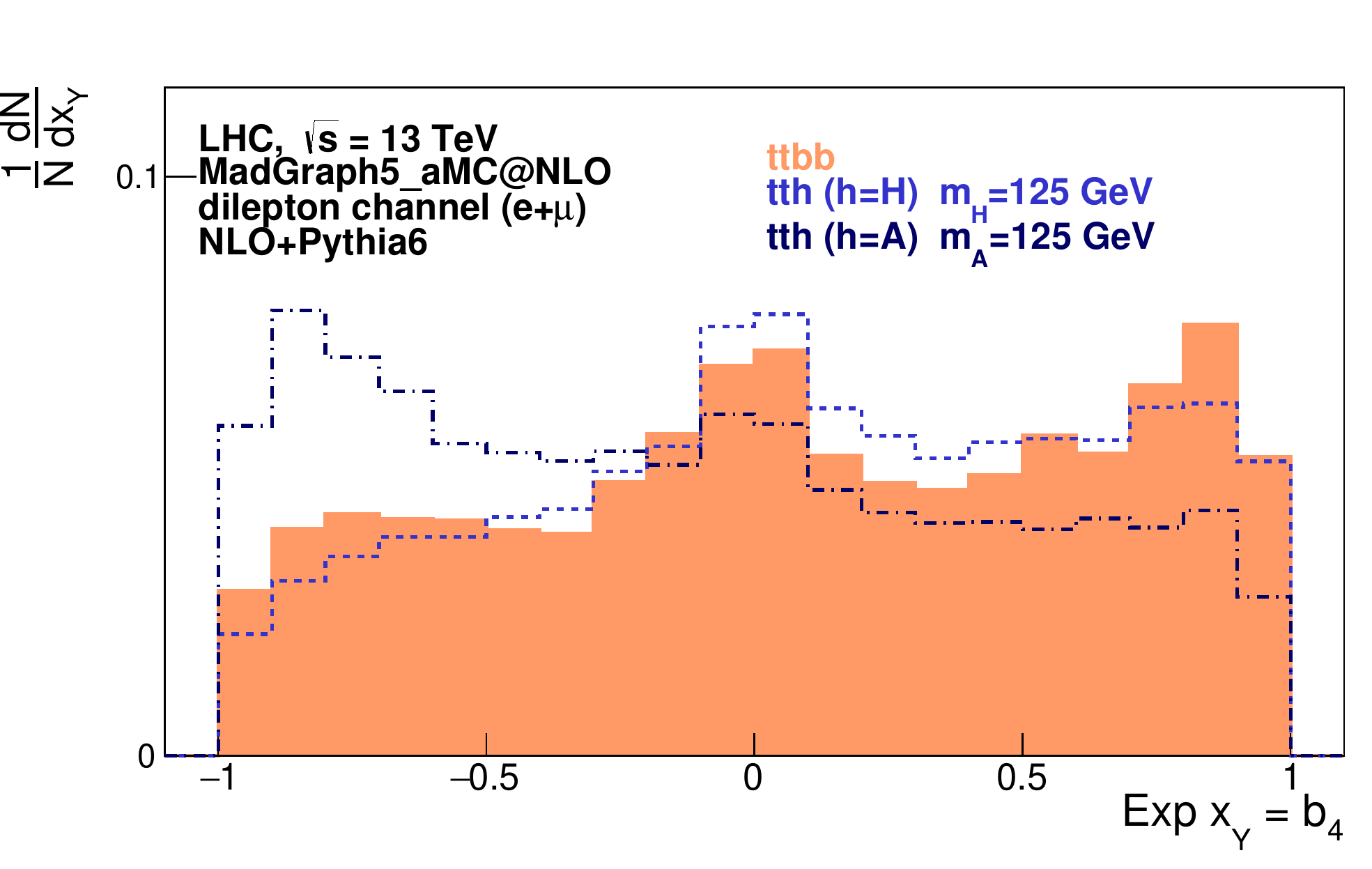} & \quad & 
\includegraphics[width=0.45\textwidth]{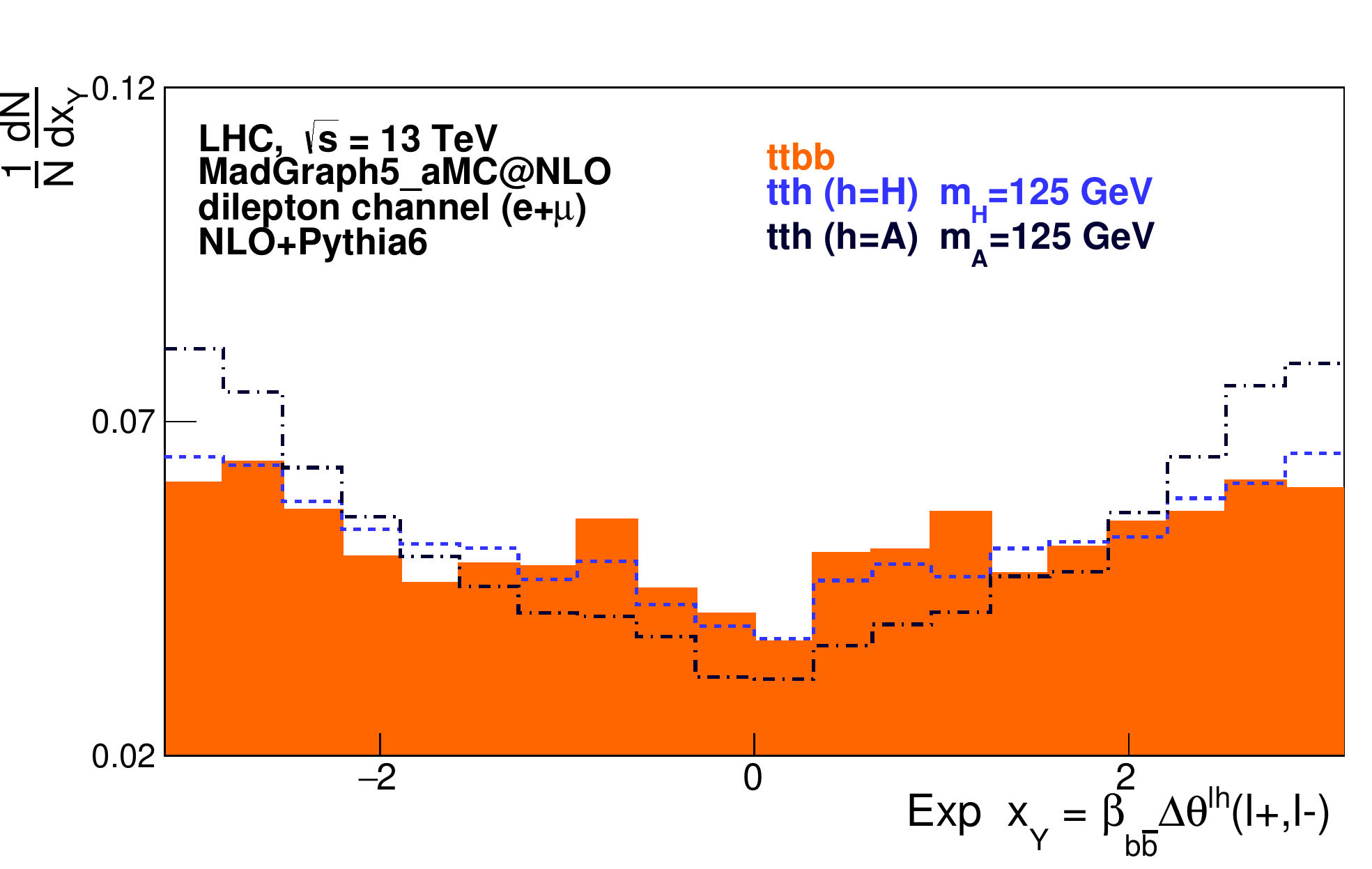}\\[-4mm]
\end{tabular}
\caption{Normalised distributions of signal events in the SM ($t\bar t H$) and pure $CP$-odd ($t\bar t A$) scenarios and of the $t\bar t b \bar b$ background, after cuts and reconstruction.}
\label{dists}
\end{center}
\end{figure}

The angular observables introduced in \cite{angdist} are constructed from angles $\theta^X_Y$, defined as the angle between the direction of $Y$, in the rest frame of $X$, and the direction of $X$ in the rest frame of its parent system. In Figure \ref{dists}, normalised distributions of $\sin(\theta^{t\bar t h}_t)\sin(\theta^{h}_{W^+})$ and $\sin(\theta^{t\bar t h}_h)\sin(\theta^{t\bar t}_{\bar t})$ are shown after cuts and kinematic reconstruction for the SM signal, pure $CP$-odd scenario signal and irreducible background $t\bar t b\bar b$ events. These are among the most sensitive observables of this kind with respect to the $CP$ nature of the signal.

From the survey of observables proposed in previous theoretical or phenomenological studies, the ones found to be more sensitive to the different $CP$ scenarios, after cuts and reconstruction, were $\beta_{b\bar b}\Delta\theta^{lh}(\ell^+,\ell^-)$
, proposed in \cite{rohini2}, and $b_4=p_t^zp_{\bar t}^z/p_tp_{\bar t}$, proposed in \cite{gunion}. Distributions of these observables are shown, after cuts and reconstruction, in Figure \ref{dists}, for signal events in the SM and in the pure $CP$-odd scenarios and for the $t\bar t b\bar b$ background.
Overall, $b_4$ is found to be the most sensitive observable, with associated ``forward-backward'' asymmetries of +0.16 for SM signal, -0.17 for pure $CP$-odd signal and 0.12 for $t\bar t b\bar b$.

\section{Expected limits on $t\bar th$ production cross-section}

\begin{figure}
\begin{center}
\begin{tabular}{ccc}
\includegraphics[width=0.5\textwidth]{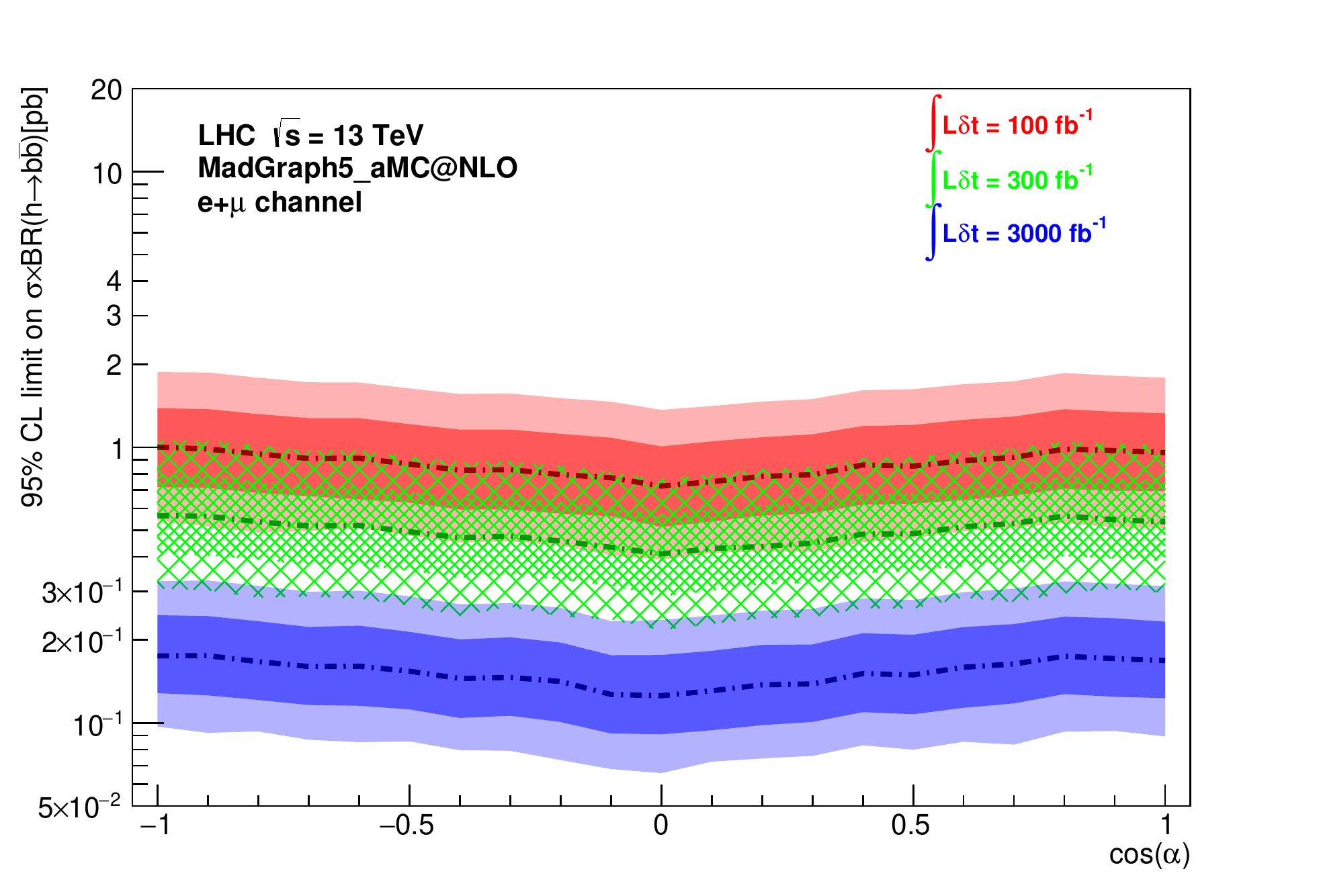} & \quad & 
\includegraphics[width=0.5\textwidth]{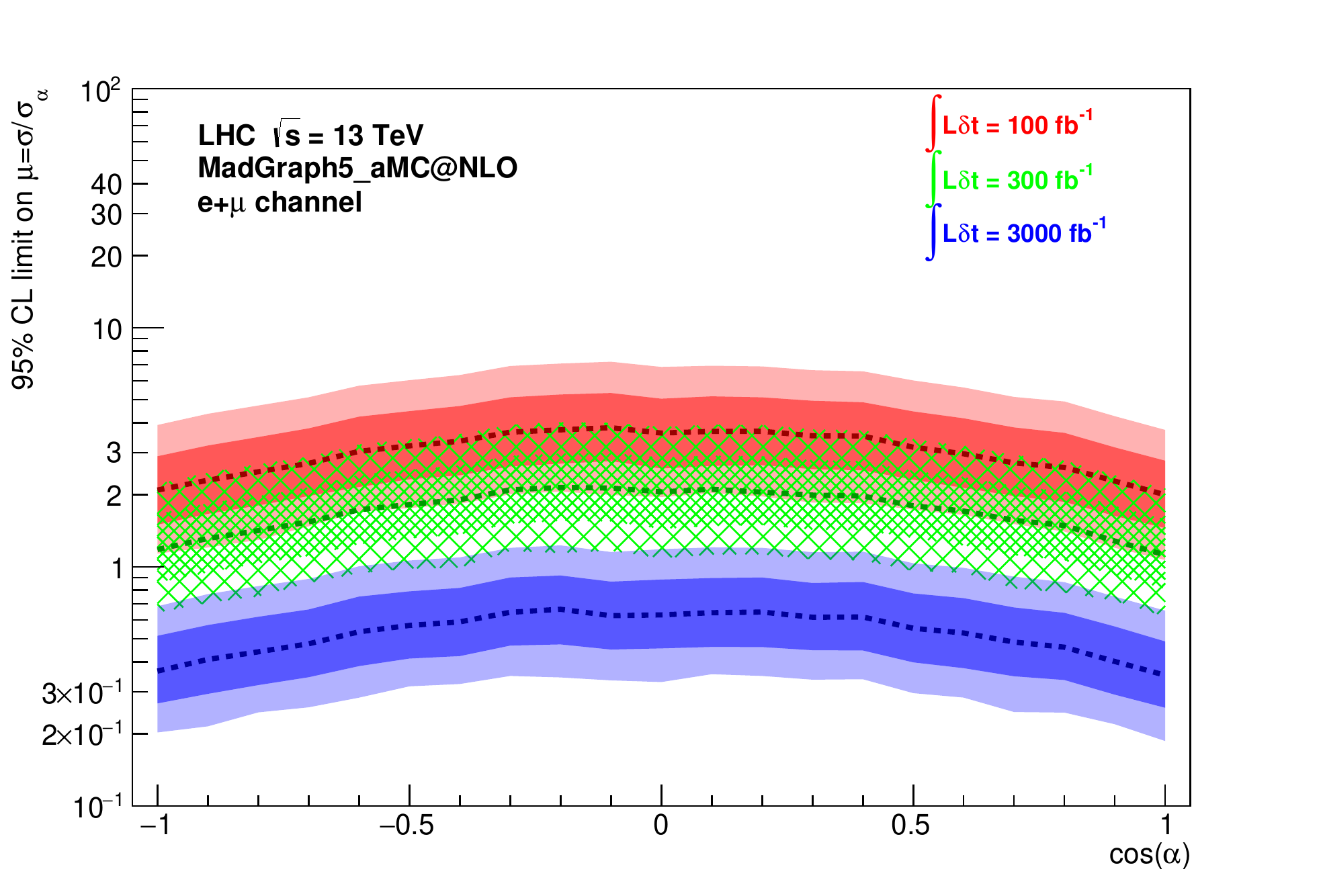}\\[-2mm]
\end{tabular}
\caption{Expected 95\% CL limits in the background-only scenario as a function of $\alpha$ and for 3 integrated luminosities. Left: $\sigma\times$BR($h\rightarrow b\bar b$); right: $\mu=\sigma / \sigma_{\alpha}$.}
\label{limits}
\end{center}
\end{figure}

A signal-background classification BDT, combining the most sensitive observables, was trained for each value of $\cos\alpha$. Its output was used as a discriminant for extracting the expected 95\% CL upper limits on the production cross-section times the $h\rightarrow b\bar b$ branching ratio ($\sigma\times\textrm{BR}$), under the background-only hypothesis. Figure \ref{limits} shows, on the left, these limits as a function of $\cos\alpha$, for 3 different integrated luminosities and, on the right, the corresponding limits on signal strength, computed relatively to the prediction for that value of $\alpha$.
From the left plot, we see that the analysis becomes more sensitive as the $CP$-odd component of the coupling increases. Thus, a search optimized for the SM $t\bar t h$ search is not insensitive to production through a mixed-$CP$ coupling. 
With respect to CP measurements, it was recently suggested that the single-lepton channel is nearly twice as sensitive as the dilepton channel \cite{duarte} and it would seem beneficial to combine the two.

\pagebreak

\Acknowledgements
I am thankful to the International Advisory Committee for granting me the opportunity to present and discuss this work at the Workshop, and to the Local Organizing Committee for a very nicely organized gathering.



\bibliographystyle{unsrt}

\begin{thebibliography}{99}
   
\bibitem{higgsdiscoveryatlas}
  G.~Aad {\it et al.} [ATLAS Collaboration],
  Phys.\ Lett.\ B {\bf 716} (2012) 1
  
\bibitem{higgsdiscoverycms}
  S.~Chatrchyan {\it et al.} [CMS Collaboration],
  Phys.\ Lett.\ B {\bf 716} (2012) 30
  
\bibitem{madspin}
  P.~Artoisenet {\it et al.},
  JHEP {\bf 1303} (2013) 015
  
\bibitem{rohini2}
  F.~Boudjema {\it et al.},
  Phys.\ Rev.\ D {\bf 92} (2015) no.1,  015019
  
\bibitem{gunion}
  J.~F.~Gunion and X.~G.~He,
  Phys.\ Rev.\ Lett.\  {\bf 76} (1996) 4468
  
\bibitem{demartin}
  F.~Demartin {\it et al.},
  Eur.\ Phys.\ J.\ C {\bf 74} (2014) no.9,  3065
  
\bibitem{angdist}
  S.~P.~Amor dos Santos {\it et al.},
  Phys.\ Rev.\ D {\bf 92} (2015) no.3,  034021
  
\bibitem{cpangvars}
  S.~Amor Dos Santos {\it et al.},
  Phys.\ Rev.\ D {\bf 96} (2017) no.1,  013004
  
\bibitem{madgraph}
  J.~Alwall {\it et al.},
  JHEP {\bf 1407} (2014) 079
  
\bibitem{pythia}
  T.~Sjostrand, S.~Mrenna and P.~Z.~Skands,
  JHEP {\bf 0605} (2006) 026
  
\bibitem{delphes}
  J.~de Favereau {\it et al.} [DELPHES 3 Collaboration],
  JHEP {\bf 1402} (2014) 057
  
\bibitem{duarte}
  D.~Azevedo, A.~Onofre, F.~Filthaut and R.~Gonçalo,
  arXiv:1711.05292 [hep-ph].




\end{thebibliography}

\end{document}